# Exponential suppression of thermal conductance using coherent transport and heterostructures

Wah Tung Lau (留華東)[*], Jung-Tsung Shen (沈榮聰)[^], and Shanhui Fan (范汕洄)[#]

Edward L. Ginzton Laboratory, Stanford University, Stanford, California 94305, USA

Date: August 6, 2010

Abstract

We consider coherent thermal conductance through multilayer photonic crystal heterostructures, consisting of a series of cascaded non-identical photonic crystals. We show that thermal conductance can be suppressed exponentially with the number of cascaded crystals, due to the mismatch between photonic bands of all crystals in the heterostructure.

[*]wahtung@gmail.com
[^]jushen@gmail.com
[#]shanhui@stanford.edu

Coherent thermal transport, using thermal channels with length smaller than the mean free path of the thermal carriers, is important for fundamental study of thermal processes, and for new device opportunities in thermal management [1-11]. In this letter, we consider a coherent thermal channel formed by cascading a series of non-identical photonic crystals. We show that its thermal conductance can be suppressed *exponentially* with respect to the channel length $l$. This is fundamentally different from the incoherent process, where the conductance typically decreases as $1/l$ [12]. We therefore demonstrate that coherent processes can be very effective in *suppressing* thermal conductance. This result also indicates that an aperiodic coherent thermal channel is qualitatively different from all previously considered coherent thermal channels [1-10] including periodic multilayer photonic crystals as we previously considered [9,10], where the intrinsic thermal conductance of the channels were all independent of the channel length $l$.

As a concrete implementation, we utilize the concept of photonic crystal heterostructure [13-15], and consider multilayer systems illustrated in Fig. 1(a), consisting of a total of $N$ different photonic crystals. All crystals consist of alternate periodic layers of vacuum and dielectric. The use of vacuum ensures that heat transfer is carried only by photons. The dielectric can be silicon ($n = n_{Si} = 3.42$). For photon frequencies within the blackbody spectrum at room temperature, silicon has very little dispersion and dissipation, with typical attenuation length exceeding millimeters [16]. Consequently photonic thermal transport should be coherent across the entire structure [9,10].

For the structure shown in Fig. 1(a), we assume that all crystals have the same period $a$. The $m^{th}$ crystal has a dielectric layer thickness $d_{sm}$ and a vacuum layer thickness $d_{vm}$. At either ends of the structure, we have two semi-infinite photonic crystals. In between these two ends, there can be a series of crystals cascaded together, each having the same number of periods $N_P$ and hence the same total thickness. (Thus the length of the channel is proportional to $N-2$.) We also assume that the two crystals at the ends are maintained at temperatures of $T$ and $T+dT$ respectively. The three-dimensional (3D) thermal conductance per unit area is then defined as $G_{3D}(T) = dQ(T)/dT$, where $dQ(T)$ is the heat flux per unit area.

In such multilayer structures, each photon state is characterized by three parameters: frequency $\omega$, wavenumber $k_{//}$ in the direction parallel to the layers, and polarization $\sigma = s, p$. The $s$ and $p$ polarizations have electric and magnetic fields parallel to the layers, respectively [17]. Summing over all photon states, we have [10]:

$$G_{3D}(T) = \sum_{\sigma} \int \frac{k_{//} dk_{//}}{2\pi} G(T, k_{//}, \sigma). \quad (1)$$

Here

$$G(T, k_{//}, \sigma) = k_B \int_0^\infty \frac{d\omega}{2\pi} \frac{[\hbar\omega/(k_B T)]^2 e^{\hbar\omega/(k_B T)}}{[e^{\hbar\omega/(k_B T)} - 1]^2} \Theta(\omega, k_{//}, \sigma), \quad (2)$$

is the thermal conductance of a one-dimensional (1D) thermal channel of a given parallel wavenumber $k_{//}$ and polarization $\sigma$. $k_B$ is the Boltzmann constant, $\hbar = h/(2\pi)$ is the reduced Planck constant. The factor $\Theta(\omega, k_{//}, \sigma)$ measures the contribution from photon states at $(\omega, k_{//}, \sigma)$. For a single crystal (i.e. $N = 1$), $\Theta(\omega, k_{//}, \sigma) = 1$ if the frequency $\omega$ lies in a photonic band, and $\Theta(\omega, k_{//}, \sigma) = 0$ if the frequency $\omega$ lies in a band gap region [10]. For the heterostructure [Fig. 1(a)], where there are at least two crystals, (i.e. $N \geq 2$), $\Theta(\omega, k_{//}, \sigma)$ vanishes unless, at a given $(\omega, k_{//}\sigma)$, propagating eigenmodes exist for the semi-infinite crystals at both ends. In such a case, $\Theta(\omega, k_{//}, \sigma)$ is the power transmission coefficient through the entire structure when an incident wave is one of these eigenmodes, and can be determined using the transfer matrix method [17]. We will first consider the behaviors of one-dimensional conductance at normal incidence, and suppress the labels $k_{//}$ and $\sigma$.

For any one-dimensional channel consisting of homogeneous dielectric material, $\Theta(\omega) = 1$ for all frequencies. Eq. (2) then leads to the universal quantized thermal conductance $G_0(T) = \pi k_B T /(6\hbar)$ [1,2,4]. For one-dimensional channel in general, we define its normalized thermal conductance as $G(T)/G_0(T)$.

We now briefly review the property of thermal conductance suppression of a single photonic crystal at normal incidence [10]. The suppression is particularly effective at the ergodic limit of $k_B Ta/(hc) \gg 1$, where the thermal conductance has contribution from a large number of photonic bands [10]. In this limit, the statistical properties of the photonic band structures as a whole, rather than the detailed properties of individual bands, become important. From Eq. (2), we thus obtain:

$$\lim_{T \to \infty} \frac{G(T)}{G_0(T)} = \lim_{\Lambda \to \infty} \frac{1}{\Lambda} \int_0^\Lambda d\omega \Theta(\omega) \equiv \eta, \text{ for } N = 1. \quad (3)$$

Here $\eta < 1$ denotes the proportion of frequencies in the photonic bands. Using ergodic theory, we showed in Ref. [10] that except for a zero-measure set of parameter choices, all crystals satisfy:

$$\eta = \int_0^\pi d\phi \{\text{Re}[\cos^{-1} \frac{(\xi+1)\cos\phi - 1}{\xi - 1}] - \text{Re}[\cos^{-1} \frac{(\xi+1)\cos\phi + 1}{\xi - 1}]\}, \quad (4)$$

where $\xi = 0.5(n + 1/n)$. We emphasize that $\eta$, and hence the ergodic limit of the normalized conductance, is independent of the layer thicknesses in the unit cell.

In a heterostructure with $N$ crystals [Fig. 1(a)], photonic band gaps are located at different frequencies for different crystals [Fig. 1(b)]. To determine its thermal conductance in the ergodic limit, we first assume that thermal transport occurs only at frequencies that fall within the bands of *all* crystals, and set $\Theta(\omega) = 0$ for all other frequencies. We then assume that the frequency distributions of the photonic bands between different crystals are independent. From these two assumptions, the total proportion of frequencies that contributes to thermal conductance is then $\eta^N$, with $\eta$ given by Eq. (4). Finally, for each frequency $\omega$ that does contribute, the power transmission coefficient $\Theta(\omega)$ is typically less than unity, due to impedance mismatch between different crystals. Combining all these considerations, we have:

$$\lim_{T \to \infty} \frac{G(T)}{G_0(T)} = \lim_{\Lambda \to \infty} \frac{1}{\Lambda} \int_0^\Lambda d\omega \Theta(\omega) < \eta^N, \text{ for } N \geq 2. \quad (5)$$

Eq. (5) is the main result of this paper: The thermal conductance of a heterostructure in the ergodic limit has an upper bound $\eta^N$ that decreases exponentially with the number of crystals $N$. Moreover, such an upper bound is independent of the detailed geometry such as layer thickness and lattice constant, but instead depends only on the refractive indices of the layers.

We now support the theoretical analysis above with extensive numerical simulations. We first verify that the ergodic limit indeed exists. In Fig. 2(a), we plot the temperature dependence of $G(T)/G_0(T)$ for various heterostructures. $G(T)/G_0(T)$ all decrease rapidly at small $T$ and converges to specific values when $T \gg hc/(k_B a)$. These specific values: $\lim_{T \to \infty}[G(T)/G_0(T)]$, which we refer as the "high-$T$ conductance", define the ergodic limit for the conductance in a given structure. Below, we will only consider conductance at this ergodic limit.

In a heterostructure, except for the two crystals at the ends, all intermediate crystals have a finite number of periods $N_P$. At frequencies in the band gaps of any intermediate crystal, photons can still transmit through the crystal, resulting in a non-zero $\Theta(\omega)$. Such a finite-size effect, however, should be negligible if $N_P$ is sufficiently large. In Fig. 2(b), we consider heterostructures with different number of crystals $N$. For each $N$, we vary $N_P$. The high-$T$ conductance all converges when $N_P \geq 10$. Thus, with a choice of $N_P = 10$, all intermediate crystals can essentially be treated as infinite, and thus $\Theta(\omega) = 0$ unless at the frequency $\omega$ there are photon states in the band structures for all crystals, validating the first assumption made in our analysis. Regarding the second assumption, one can in fact prove that frequency distributions of photon states in different crystals are independent, except for a zero-measure set of parameter choices, using an analysis similar to Ref. [10].

Eq. (5) predicts that an *upper bound* that is universal in the sense that it is independent of layer thicknesses. Numerical calculations establish an even stronger result. In Fig. 2(c), we consider a total of 15,000 structures, with the number of crystals $N$ between 2 and 4. In these structures, the thickness of the dielectric layers is randomly generated. For the vast majority of structures, the high-$T$ conductance themselves have universal values that depend only on $N$ and are independent of the layer thicknesses. This numerical result provides a strong support of our statistical analysis that leads to Eq. (5).

Examining Fig. 2(c), we also see few cases for which the high-$T$ conductance of a structure deviates from the universal values. [One of these cases is also shown in Fig. 2(a).] Such deviations occur if one of the crystals belongs to the commensurate cases either as discussed in Ref. [10], or when photonic bands of two or more crystals become correlated to each other. These cases, in the ergodic limit, form the zero-measure set as discussed above. We will ignore these commensurate cases hereafter, since their occurrences require control of layer thicknesses with high accuracy [10].

We now provide a direct numerical confirmation of Eq. (5). In Fig. 3, we plot the high-$T$ conductance of heterostructures with different number of crystals $N$. We consider only the incommensurate cases where the high-$T$ conductance takes the universal values. The numerically determined high-$T$ conductance clearly falls below the exponentially-decreasing bound as set by Eq. (5).

For three-dimensional thermal conductance, using Eqs. (1) and (2), we have [10]:

$$G_{3D}(T) = \sum_{\sigma=s,p} \int_0^n u\,du \int_0^\infty d\omega \frac{\omega^2}{4\pi^2 c^2} \{k_B \frac{[\hbar\omega/(k_B T)]^2 e^{\hbar\omega/(k_B T)}}{[e^{\hbar\omega/(k_B T)}-1]^2} \Theta(\omega,u,\sigma)\}, \qquad (6)$$

where $u \equiv k_\parallel/\omega$. At the ergodic limit, for each channel labeled by $u$, we repeat the same statistical arguments above, to obtain:

$$\lim_{T\to\infty} \frac{G_{3D}(T)}{G_{vac}(T)} = \sum_{\sigma=s,p} \int_0^1 u\,du \cdot \lim_{\Lambda\to\infty} \frac{1}{\Lambda}\int_0^\Lambda d\omega \Theta(\omega,u,\sigma)$$
$$\leq \sum_{\sigma=s,p} \int_0^1 [\eta(u,\sigma)]^N u\,du. \qquad (7)$$

Here $\eta(u,\sigma)$ is obtained by replacing $\xi$ in Eq. (4) with $\xi(u,\sigma) = 0.5[n_a/(n_b P) + (n_b P)/n_a]$, where $n_a = \sqrt{n^2 - u^2}$, $n_b = \sqrt{1-u^2}$, and $P=1$ and $n^2$ for the $s$ and $p$ polarizations, respectively.

In Eq. (7), the equality holds only when $N=1$. Also, in deriving Eq. (7), we use the fact that in the ergodic limit, the states with $u > 1$, where photons are evanescent in the vacuum layers, do not contribute [10]. Hence the range of integration on $u$ is between 0 and 1. Finally, Eq. (7) indicates a universal upper bound that depends only on the index and the number of crystals. Similar to the case of normal incidence, numerical results (not shown here) further indicate that 3D conductance themselves in the ergodic limit are in fact universal.

We now confirm Eq. (7) with numerical results. For the *s* polarization, the band gaps persist in the entire range of $0 < u < 1$ [9,10]. Hence $\eta(u, \sigma = s) < 1$. From Eq. (7), the upper bound on the high-*T* conductance for the *s* polarization therefore decreases exponentially with respect to the number of crystals $N$, as supported by the numerical results in Fig. 4. For the *p* polarization, $u_B = \sqrt{n^2/(n^2+1)}$ corresponds to the Brewster angle where the reflection at the vacuum-dielectric interface vanishes [18,19]. At $u = u_B$ there is no photonic band gap at any frequency, and $\eta(u_B, \sigma = p) = 1$. As a result, the high-*T* conductance from the *p* polarization, and hence the total 3D high-*T* conductance, show a more gradual decay with respect to $N$ (Fig. 4).

In our structure, exponential reduction of 3D thermal conductance can be accomplished using a simple angle filter that blocks out thermal radiation with a large angle of incidence. As a simple illustration, restricting the integration range of $u$ in Eq. (6) to $0 \leq u \leq 0.5$ is sufficient to generate the exponential reduction (Fig. 4). Alternatively, for heterostructures between two semi-infinite vacuum regions, we can achieve exponential reduction using omnidirectional reflectors as the intermediate crystals [18]. Finally, Ref. [20] showed that the Brewster angle can be made imaginary using films with large birefringence. Multilayer structures consisting of such birefringent films and vacuum therefore could also be useful for demonstrating exponential reduction of 3D thermal conductance.

We now briefly comment on the experimental feasibility in demonstrating our predictions. Multilayer structures can be fabricated by a variety of techniques [18,20]. Our theory calls for a measurement of the photonic thermal conductance in these structures. For this purpose we notice the exciting recent experimental developments in measuring photonic thermal conductance [21,22]. Here, one can consider a set up similar to [21,22], where thermal contacts are made to the ends of the structure, and thermal conductance is measured by recording the amount of power transferred. Our results may also be practically significant for ultimate thermal insulation. Currently, an effective mechanism for thermal insulation utilizes multiple layers of metal and vacuum [23,24]. In such a structure, thermal transfer is incoherent. Its thermal conductance scales as $1/N_M$, where $N_M$ is the number of metal layers [23,24]. Our results here indicate a qualitatively different mechanism for achieving thermal insulation using dielectric systems, which may present new opportunities. For example, the structure in Fig. 1 allows high transmission through narrow band of frequencies, which may be important for communication through such a thermally insulating medium.

We end by highlighting closely related works. The concept of using a single phononic crystal structure to suppress coherent phononic thermal conductance was discussed in Ref. [25], and was extended to the photonic case in Refs. [9] and [10]. In these structures, the thermal conductance is independent of the channel length. Mismatch between two different phononic band structures had been used to achieve very small interfacial thermal conductance [26,27]. Our result, showing exponential suppression of thermal conductance using a large number of crystals, is a step further and points to a new regime of coherent thermal transport. While we have considered photonic thermal transport, the general concept in this paper may be applicable to other thermal carriers as well provided that the coherent effect is significant. Finally, in analogy to electron transport, where an exponential reduction of electronic conductance with respect to distance is a direct signature of electron localization, our result indicates an intriguing notion that heat can be localized as well.

This work is supported by AFOSR-MURI programs (Grants No. FA9550-08-1-0407 and No. FA9550-09-1-0704), and by AFRL.

**References**
[1] A. Greiner *et al.*, Phys. Rev. Lett. **78**, 1114 (1997).
[2] L. G. C. Rego *et al.*, Phys. Rev. Lett. **81**, 232 (1998).
[3] M. V. Simkin *et al.*, Phys. Rev. Lett. **84**, 927 (2000).
[4] K. Schwab *et al.*, Nature (London) **404**, 974 (2000).
[5] D. G. Cahill *et al.*, J. Appl. Phys. **93**, 793 (2003).
[6] T. Yamamoto *et al.*, Phys. Rev. Lett. **92**, 075502 (2004).
[7] D. R. Schmidt *et al.*, Phys. Rev. Lett. **93**, 045901 (2004).
[8] M. Meschke *et al.*, Nature (London) **444**, 187 (2006).
[9] W. T. Lau *et al.*, Appl. Phys. Lett. **92**, 103106 (2008).
[10] W. T. Lau *et al.*, Phys. Rev. B **80**, 155135 (2009).

[11] W. Kim *et al*., Nanotoday **2**, 40 (2007).

[12] C. Kittel and H. Kroemer, *Thermal Physics*, (W. H. Freeman, 1980).

[13] X. Wang *et al*., Appl. Phys. Lett. **80,** 4291 (2002).

[14] B. S. Song, *et al*., Science **300**, 1537 (2003).

[15] E. Istrate *et al*., Rev. Mod. Phys. **78**, 455 (2006).

[16] E. D. Palik, Handbook of Optical Constants of Solids (Academic, New York, 1985), p. 554.

[17] A. Yariv and P. Yeh, *Photonics: Optical Electronics in Modern Communications*, 6th ed. (Oxford University Press, Oxford, England, 2007).

[18] Y. Fink *et al*., Science **282**, 1679 (1998).

[19] J. D. Jackson, *Classical Electrodynamics*, (Wiley 1998).

[20] M. F. Weber *et al*., Science **287**, 2451 (2000).

[21] S. Shen *et al*., Nano Lett. **9,** 2909 (2009).

[22] E. Rousseau *et al*., Nature Photonics **3,** 514 (2009).

[23] G. R. Cunnington *et al*., Progress in Astronautics and Aeronautics **23,** 111 (1970).

[24] C. L. Tien *et al*., Adv. in Heat Transfer **9,** 349 (1973).

[25] G. Chen, ASME J. Heat Transfer **121**, 945 (1999).

[26] V. Narayanamurti *et al*., Phys. Rev. Lett. **43**, 2012 (1979).

[27] H. K. Lyeo *et al*., Phys. Rev. B **73**, 144301 (2006).

Figure 1

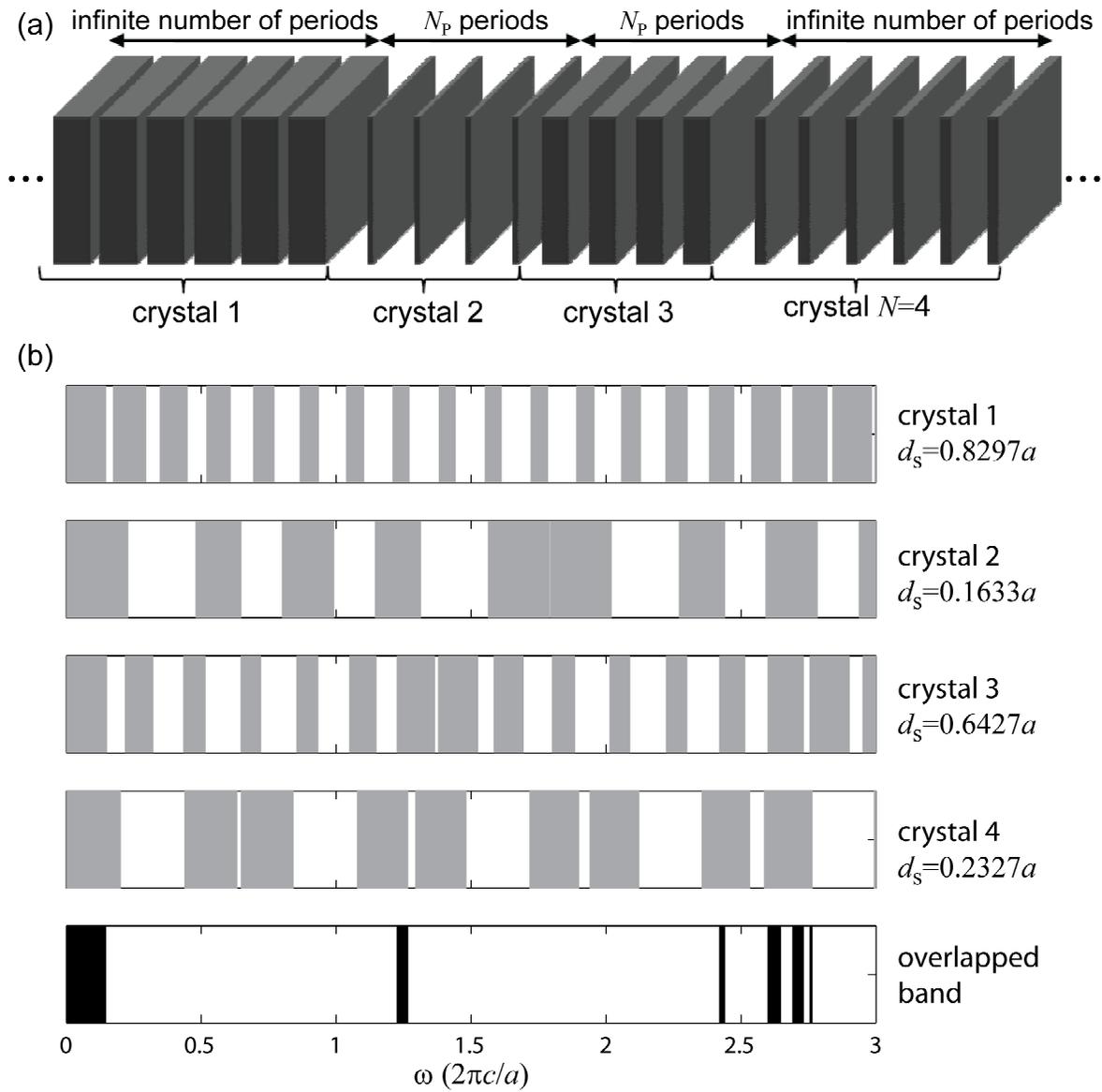

FIG. 1. (a) A photonic crystal heterostructure with 4 crystals. Each crystal has alternate silicon and vacuum layers, and a period $a$. The two crystals at either ends are assumed to be infinite.

(b) In the top 4 panels, gray regions correspond to frequency ranges where the corresponding crystal, assumed to be infinite, has a propagating state along the normal incident direction. White regions are the band gaps. $d_s$ is the thickness of the silicon layers. In the bottom panel, the black regions represent frequency ranges where there are photonic bands in all crystals.

Figure 2

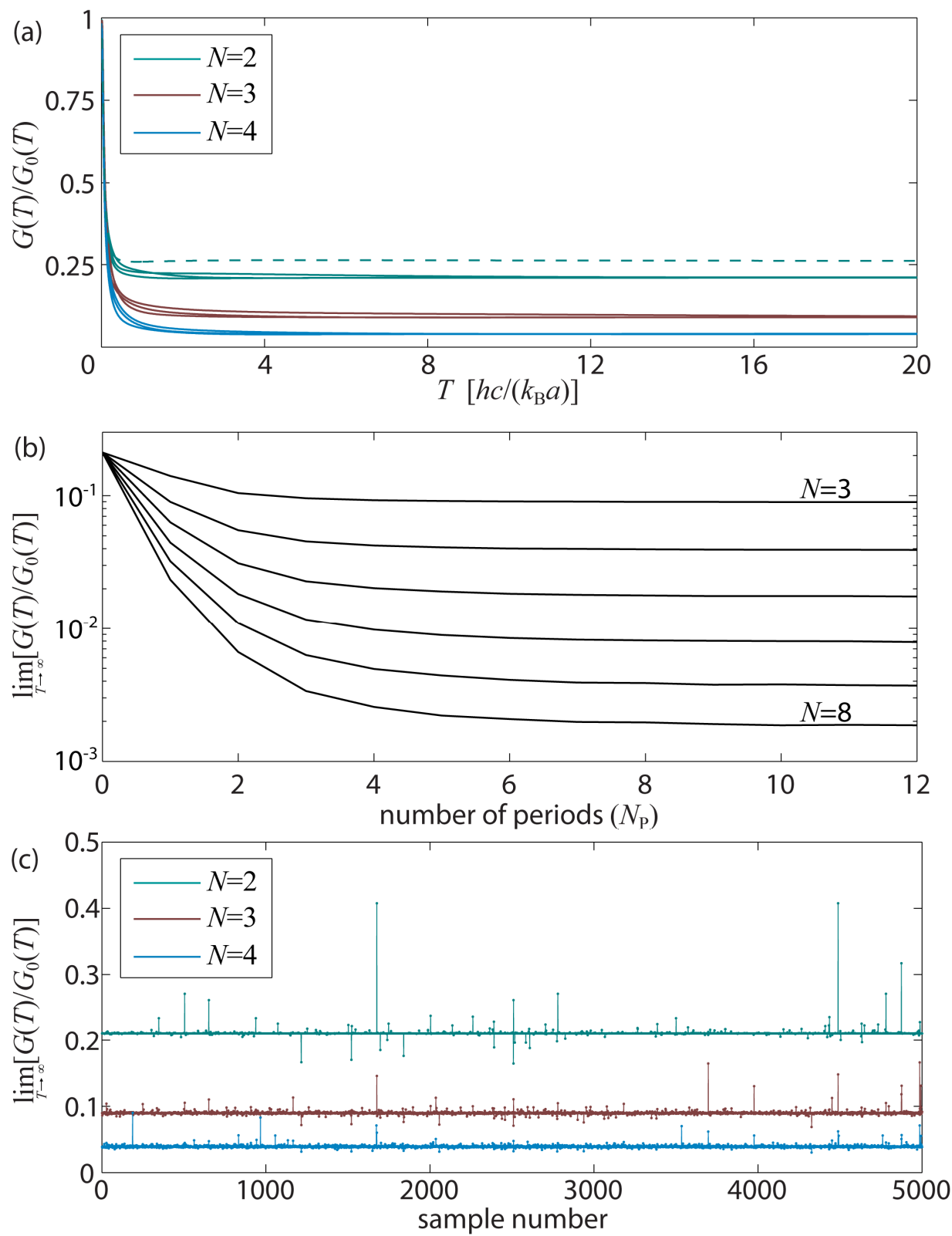

FIG. 2. (Color online) (a) Normalized one-dimensional thermal conductance $G(T)/G_0(T)$ as a function of temperature $T$ for heterostructures with different number of crystals $N$. For each $N$, different curves represent structures of different thicknesses of silicon $d_s$ and vacuum $d_v$. The green dashed curve shows a commensurate case of $N=2$, where $d_{s1} = 0.7851a$ and $d_{s2} = 0.2262a$ such that $n_{si} d_{s2}/(a-d_{s2}) = 1$.

(b) High-$T$ conductance for heterostructures of different number of periods $N_P$ at the intermediate crystals. Each line corresponds to structures with the same number of crystals $N$. $3 \leq N \leq 8$. Data points exist only at integer $N_P$.

(c) High-$T$ conductance for a total of 15,000 heterostructures with randomly generated thicknesses of silicon layers. $0.05a \leq d_s \leq 0.95a$ for all crystals.

Figure 3

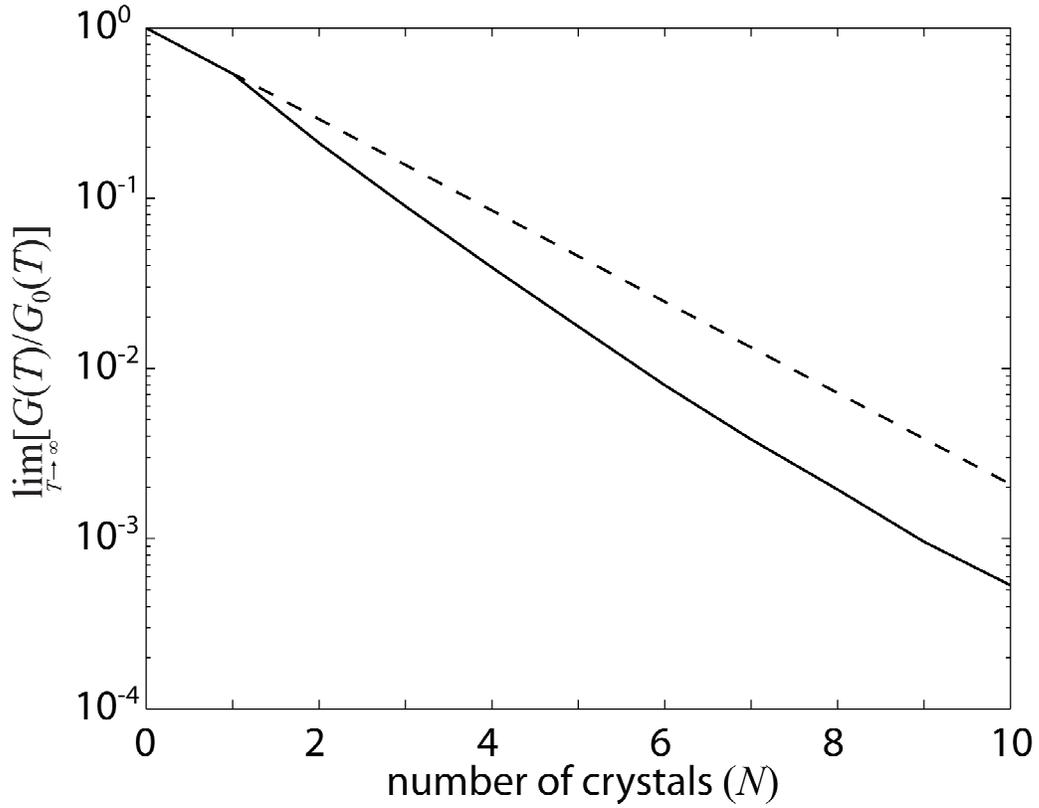

FIG. 3. One-dimensional high-$T$ conductance for photonic crystal heterostructures as a function of number of crystals ($N$). Data points exist only at integer $N$. $N=0$ corresponds to vacuum. $N=1$ corresponds to an infinite crystal. $N=2$ corresponds to two semi-infinite crystals placed together. For $N \geq 3$, each intermediate crystal has $N_P = 10$ number of periods. The solid line is the actual conductance. The dashed line is the corresponding theoretical upper bound $\eta^N \equiv \exp(-\alpha N)$ from Eq. (5), where $\alpha = 0.62$.

Figure 4

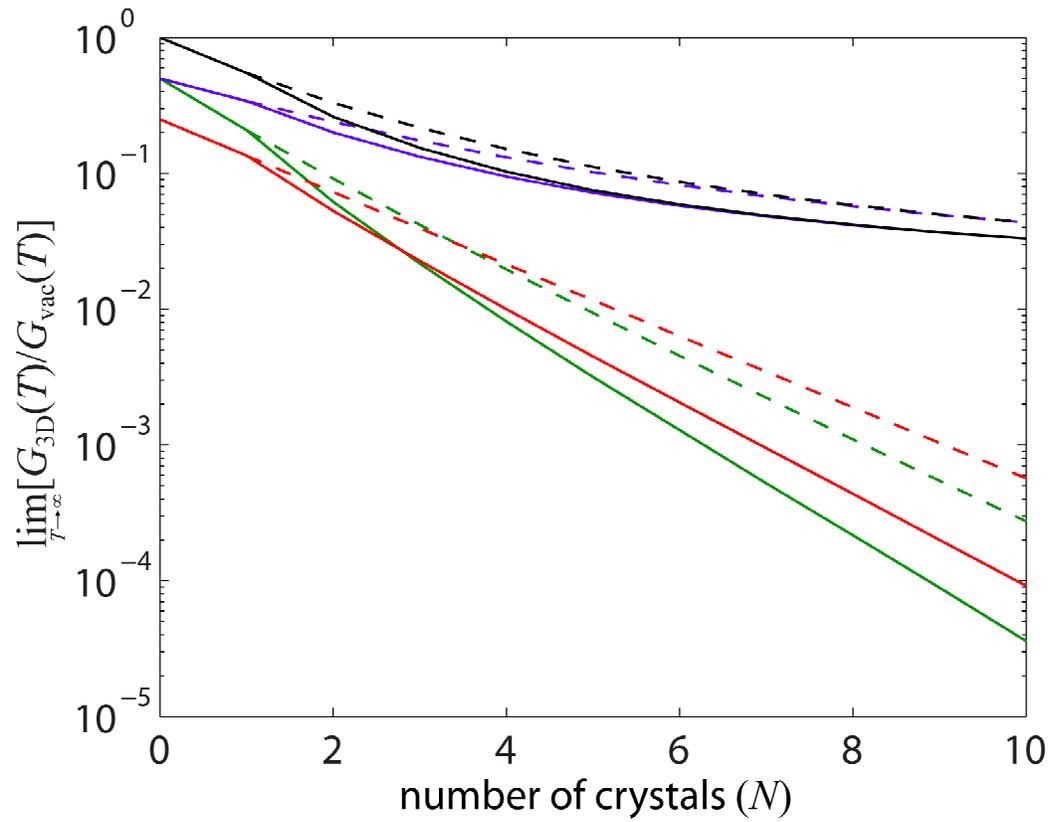

FIG. 4. (Color online) 3D high-$T$ thermal conductance versus number of crystals ($N$) in the structures. Data points exist only at integer $N$. The black lines are the total conductance. The red lines are the total conductance with an angle filter. The green and the purple lines are contributions from the $s$ and $p$ polarizations respectively. The solid lines are direct numerical calculations of the thermal conductance. The dashed lines are the theoretical upper bound from Eq. (7). For $N \geq 3$, each intermediate crystal has $N_\text{P} = 10$ number of periods.